\documentclass[11pt]{article}
\usepackage{hyperref}
\pdfoutput=1
\usepackage{hyperref}
 \pdfoutput=1
\begin{document}
\title{Attractive internal wave patterns}
\author{Jeroen Hazewinkel$^{1,2,\#}$ and Leo R.M. Maas$^{1}$ and Stuart B. Dalziel$^{2}$\\
\\\vspace{6pt} 1) Royal Netherlands Institute for Sea Research, the Netherlands\\
2) Department of Applied Mathematics and Theoretical Physics,\\ 
University of Cambridge, UK\\
\# presently at: Scripps Institute of Oceanography, La Jolla, USA
}
\maketitle
\begin{abstract}
This paper gives background information for the fluid dynamics video on internal wave motion in a trapezoidal tank. 
\end{abstract}

\noindent  Continously statified fluids that are subjected to a periodic forcing can reveal the existence of internal wave attractors \textit{Maas et al. (1997), Hazewinkel et al.( 2008), Hazewinkel et al. (2010))}. These attractors reveal themselves as those locations in the stratified fluids where all internal wave energy is found and subsequently the highest amplitudes of the wave are found. In a uniformly stratified  fluids this is most easily seen. The dispersion relation for internal gravity waves in such a fluid relates the frequency $\omega=N\cos\theta$ to buoyancy frequency $N$ and energy propagation angle $\theta$. All fluids are confined to a certain domain and ray tracing the energy shows the existence of attractors in most confined fluids \textit{Maas and Lam (1995)}. Experiments in trapezoidal tanks have confirmed the theoretical predictions and have focussed on the shape of wave number spectra or the robustness to perturbations. In the movie presented, the continuous stratification of the previous experiments has been replaced by a number of homogeneous layers of fluids of increasing density in depth. This simple change now allows us to see the internal waves in the fluid by means of simple shadowgraphy. The dynamics is identical to the uniformly stratified fluids studied previously.

The movie shows the wave motion found when a trapezoidal tank filled with a stratified fluid is subjected to a horizontal oscillation. The stratification consists of 18 homogeneous layers over a total depth of 26 cm with a density increase $\delta \rho=9\  kg/m^3$ per layer. The total difference in density and depth of the fluid correspond to a linearly stratified fluid with buoyancy frequency $N=2.5\ rad/s$, the default attractor studied by \textit{Hazewinkel et al. (2010))}. The forcing frequency $2\pi/5.0\ rad/s$. For a forcing amplitude of $2 cm$ a steady state attractor is found after 25 periods of oscillation, very similar to the evolution described by \textit{Hazewinkel et al.(2008)}. For an amplitude of $3.5\ cm$, the energy in the wave beams becomes too high and the waves start to overturn. 
\bibliographystyle{~/Dropbox/POF/Final/pf.bst}

\begin{thebibliography}{4}
\providecommand{\natexlab}[1]{#1}
\providecommand{\url}[1]{\texttt{#1}}
\providecommand{\urlprefix}{URL }
\expandafter\ifx\csname urlstyle\endcsname\relax
  \providecommand{\doi}[1]{doi:\discretionary{}{}{}#1}\else
  \providecommand{\doi}{doi:\discretionary{}{}{}\begingroup
  \urlstyle{rm}\Url}\fi
\providecommand{\eprint}[2][]{\url{#2}}

\bibitem[{\textit{Maas and Lam} (1995)}]{ML95}
Maas, L. R.~M., and F.-P.~A. Lam (1995), Geometric focusing of internal waves,
  \textit{Journal of Fluid Mechanics}, \textit{300}, 1--41.

\bibitem[{\textit{Maas et~al.}(1997) }]{maasetal}
Maas, L. R.~M., D.~Benielli, J.~Sommeria, and F.-P.~A. Lam (1997), Observation
  of an internal wave attractor in a confined stably-stratified fluid,
  \textit{Nature}, \textit{388}, 557--561.

\bibitem[{\textit{Hazewinkel et~al.}(2008)}]{H}
Hazewinkel, J., P.~v.~Breevoort, S.~B. Dalziel, and L.~R.~M. Maas (2008),
  Observations on the wavenumber spectrum and decay of an internal wave
  attractor, \textit{Journal of Fluid Mechanics}, \textit{598}, 373--382.


\bibitem[{\textit{Hazewinkel et~al.}(2010)}]{HTMD10}
Hazewinkel, J., C.~Tsimitri, L.~R.~M. Maas, and S.~B. Dalziel
  (2010{\natexlab{c}}), Observations on the robustness of internal wave
  attractors to perturbations, \textit{Physics of Fluids, in press}.

\end{thebibliography}

\end{document}